\documentclass[twocolumn,english,showpacs,amssymb,preprintnumbers,prc,floatfix]{revtex4}
\usepackage[T1]{fontenc}
\usepackage[latin1]{inputenc}
\usepackage{ae}
\usepackage{aecompl}
\usepackage{graphicx}
\usepackage{setspace}

\makeatletter

\providecommand{\tabularnewline}{\\}

\usepackage{rotating,epsfig,dcolumn}
\usepackage{graphicx,bm}
\def\hw{$\hbar\omega\,$}

\usepackage{babel}
\makeatother
\begin{document}

\title{Coexistence of spherical states with deformed and superdeformed bands
in doubly magic $^{40}$Ca; A shell model challenge.}

\author{E. Caurier $^{*}$, J.~Men\'{e}ndez $^{+}$, F. Nowacki $^{*}$
and A.~Poves $^{+}$}

\affiliation{({*}) IReS, B\^{a}t27, IN2P3-CNRS/Universit\'{e} Louis Pasteur
BP 28, F-67037 Strasbourg Cedex 2, France\\
 (+) Departamento de Fisica Te\'{o}rica, C-XI. Universidad Aut\'{o}noma
de Madrid, E-28049, Madrid, Spain}

\date{\today{}}

\begin{abstract}
Large scale shell model calculations, with dimensions reaching 10$^{9}$,
are carried out to describe the recently observed deformed (ND) and
superdeformed (SD) bands based on the first and second excited 0$^{+}$
states of $^{40}$Ca at 3.35~MeV and 5.21~MeV respectively. A valence
space comprising two major oscillator shells, $sd$ and $pf$, can
accommodate most of the relevant degrees of freedom of this problem.
The ND band is dominated by configurations with four particles promoted
to the $pf$-shell (4p-4h in short). The SD band by 8p-8h configurations.
The ground state of $^{40}$Ca is strongly correlated, but the closed shell
still amounts to 65\%. The energies of the bands are very well reproduced
by the calculations. The out-band transitions connecting the SD band
with other states are very small and depend on the details of the mixing
among the different np-nh configurations, in spite of that,
the calculation describes them
reasonably. For the in-band transition probabilities
along the SD band, we predict a fairly constant transition quadrupole moment
Q$_0$(t)$\sim$170~e~fm$^2$ up to J=10, that decreases toward the higher
spins. We submit also that the J=8 states of the deformed and superdeformed
band are maximally mixed.
\end{abstract}

\pacs{21.10.Sf, 21.60.Cs, 23.20.Lv, 27.40.+z, 29.30.-h}

\maketitle

\section{Introduction}
\label{sec:intro}

$^{40}$Ca is a textbook example of doubly magic nucleus, corresponding
to neutrons and protons filling the three lowest harmonic oscillator 
shells with principal quantum numbers p=0, 1, and 2. The N=Z=20 gap
is large (about 7~MeV) and therefore, in a naive independent
particle description, its spectrum will consist of negative parity 
quasi degenerated states starting at an excitation energy commensurable
with the value of the gap and positive parity states  at
about twice the value of the gap. Indeed, this is not the case and
its first excited state is a  0$^{+}$ at 3.35~MeV, that turns out to be
the head of a deformed band.
Excited deformed bands in spherical nuclei provide a spectacular example
of coexistence of very different structures at the same energy scale,
that is a rather peculiar aspect of atomic nuclei. Other cases have
been known for a long time, for instance, the four particle four holes
and eight particle eight holes states in $^{16}$O, starting at 6.05~MeV
and 16.75~MeV of excitation energy \cite{O16a,O16b}. 
The theoretical descriptions based in multiple particle hole excitations
that can accommodate deformation started with the work of
Brown and Green \cite{bg}, Gerace and Green \cite{gg1,gg2} and Zuker, Buck and McGrory \cite{zbm}
In $^{40}$Ca, there has been since long  experimental indications showing that
the two low-lying
sequences,  0$^+$,  2$^+$,  4$^+$,  starting at 3.35~MeV and 5.21~MeV, may
correspond to deformed or superdeformed bands (see ref. \cite{middle}).
However, it
is only recently that such bands, deformed and superdeformed,
have been explored up to high spins,  in several medium-light nuclei such as $^{56}$Ni
\cite{Ni56exp}, $^{36}$Ar \cite{Ar36exp} and of course $^{40}$Ca \cite{Ca40exp},
 thanks to the availability of large arrays of
$\gamma$ detectors like Euroball and Gammasphere.

One characteristic feature of these bands is that they may belong to rather
well defined  shell model configurations; for instance, the
deformed excited band in $^{56}$Ni can be associated with the configuration
(0f$_{7/2}$)$^{12}$ (1p$_{3/2}$, 0f$_{5/2}$, 1p$_{1/2}$)$^{4}$
while the super-deformed band in $^{36}$Ar is based in the structure
$(sd)^{16}$ $(pf)^{4}$. The location of the np-nh states in $^{40}$Ca
was studied in the Hartree-Fock approximation with blocked particles
and Skyrme forces in ref.~\cite{Zamick}. While many approaches are
available for the microscopic description of these bands (Cranked
Nilsson-Strutinsky \cite{Ar36exp}, Hartree-Fock plus BCS with configuration
mixing \cite{Bender}, Angular Momentum Projected Generator Coordinate
Method \cite{Rayner}, Projected Shell Model \cite{Sun}, Cluster
models \cite{Sakuda,kanada} etc.) the interacting shell model is, when affordable,
a prime choice. The mean field description of $N=Z$ nuclei, has
problems related to the correct treatment of the proton-neutron pairing
in its isovector and isoscalar channels, as well as the proper restoration
of the symmetries broken at the mean field level to take into account the 
pairing (particle number) and quadrupole (angular momentum) correlations.
In addition, the triaxial degrees of freedom are  seldom considered.
On the shell model side, the problems come from the size of the valence
spaces needed to accommodate the np-nh configurations.

In a recent article \cite{Ar36}, it was shown that large scale shell model
calculations within a valence space consisting on the $sd$ and $pf$
shells can describe the coexistence of the spherical ground state
and the superdeformed band of $^{36}$Ar, both the very existence
of the spherical and the superdeformed states forming a band and the
configuration mixing that permits transitions connecting them. In
the same spirit, in the present work we focus on the case of $^{40}$Ca,
which is particularly challenging since it presents two rotational
bands of different kind above the spherical ground state. The
first rotational band, starting at 3.35 MeV of excitation energy,
is generated by an intrinsic deformed state, whereas the other one, starting
at 5.21 MeV above the ground state, stems from an intrinsic
super-deformed state, i.e. bearing an axis ratio close to 2:1. 

In earlier works we have surmised that the superdeformed band corresponds to
a structure $(sd)^{16}$ $(pf)^{8}$, while the normal deformed band
presents a $(sd)^{20}$ $(pf)^{4}$ structure \cite{Ca40nph,arxivCa40}.
That is, the structure of the bands corresponds to the promotion of
four and eight particles across the Fermi level, from the $sd$ to
the $pf$ shell, that we call 4p-4h and 8p-8h configurations. However,
this description is too crude, because the physical states contain
components of different np-nh rank. The mixing should allow
for transitions connecting the superdeformed and normal deformed bands
as well as both bands with the spherical ground state, while remaining
 gentle enough not to jeopardize
the very existence of the bands. This is the aim of this work; to
explain the coexistence of spherical, deformed and superdeformed
structures at low energy in $^{40}$Ca and to understand how do their dominant
configurations mix among themselves while preserving their identity.

\section{Spherical mean field $vs$ correlations}
\label{sec:sph}

The intriguing question is how comes that these many particle-hole
configurations appear so low in excitation energy. Before answering it 
with the actual calculation, we want to advance a more qualitative
discussion based in the competition between spherical mean field and 
quadrupole and pairing correlations.

In a very recent paper \cite{rowe} Rowe {\it et  al.} have examined this issue
in  $^{16}$O, with a model consisting of
a spherical harmonic oscillator mean field plus an Elliott's
quadrupole-quadrupole interaction. In their model, the unperturbed excitation
energy of an np-nh excitation between contiguous shells is n\hw. With the
standard prescription for {\hw} the value of the $^{40}$Ca gap is largely
overestimated (12~MeV instead of the experimental 7~MeV). The quadrupole correlation
energy is taken as proportional to the Casimir operator of the lowest grade SU(3)
irrep corresponding to the np-nh configuration, $C_{SU(3)}$, with the coupling constant for
the  quadrupole-quadrupole interaction equal to {\hw}/2N$_0$, (N$_0$ being the total
number of oscillator quanta in the ground state of the nucleus of interest). 
The excitation energy of the np-nh band-head can be written as:
\begin{equation}
 \Delta E = \hbar \omega  \left( (\rm{N}(np-nh)-\rm{N}_0)- \frac{C_{SU(3)}}{4 N_0} \right) 
\end{equation}
For the deformed band --n=4, $\lambda$=12,  $\mu$=8-- $\Delta$E=1.06{\hw} 
and for the superdeformed band --n=8, $\lambda$=20,  $\mu$=12-- $\Delta$E=0.66{\hw}
($\lambda$ and $\mu$ being Elliott's labels of the irreps of SU(3))
This model is very schematic, and, as we shall discuss later, overestimates both the 
energy differences between the closed shell and the np-nh configurations at the mean field level,
and the correlation energies of the later. Thus, it is not surprising that it cannot reproduce the
experimental numbers. But it contains the right message; that there is a class of
states, related to SU(3) configurations of maximum  weight, that can gain huge amounts of
correlation energy provided the structure of the spherical mean field does not hamper
their collective build-up. In the present case, instead of the assumed SU(3) coupling scheme in
both shells, the physical situation is closer to pseudo-SU(3) \cite{psu3} in the $sd$ shell,
based in the quasi-degenerate  $s_{1/2}$-$d_{3/2}$ doublet and
quasi-SU(3) \cite{qsu3} in the $pf$ shell, originated in the $\Delta$j=2, $\Delta$l=2 pair of orbits, 
 $f_{7/2}$-$p_{3/2}$.  This means lower symmetry and therefore less correlation
energy. But this reduction is more than compensated by using a better spherical mean field, that, in 
particular, must incorporate quadratic monopole terms. These, as we shall see, reduce
notably the mean field excitation energy of the np-nh configurations \cite{az.phi}.
Furthermore, pairing correlations need to be taken into account as well.

In addition to the estimates of  the energy gains, the above SU(3)-like
models make it possible to compute the intrinsic quadrupole moments associated to the
different np-np deformed structures. Using the pseudo+quasi-SU(3) prescription we
find Q$_0$=125~e~fm$^2$ for the 4p-4h configuration and Q$_0$=180~e~fm$^2$ for the
8p-8h, consistent with what one would expect for a deformed and a superdeformed band.
Let's underline that  what we submit is that super-deformation in this region of nuclei can be achieved
readily in the space of two major oscillator shells, provided that SU(3)-like
geometries of the spherical mean field are at hand. 
The SU(3) limit gives Q$_0$=148~e~fm$^2$ and Q$_0$=226~e~fm$^2$ instead. As the quadrupole
correlation energy should vary as the square of the quadrupole moment, we can reckon that
the pseudo+quasi-SU(3) correlation energy represents 2/3 of the SU(3) limit.

\section{Valence space and effective interaction}
\label{sec:val}

As we have explained above, an adequate valence space for the study of
the coexisting bands in $^{40}$Ca consists of the $sd$ and $pf$ major
oscillator shells. Its only drawback is that the dimensions
involved (10$^{12}$) are beyond our present computing capability. A
possible way out is to close the $d_{5/2}$ orbit, that is, to work with a
virtual $^{28}$Si inert core. The quadrupole colectivity of the solutions will
be reduced by this truncation, but we have checked in $^{36}$Ar that
the effect is moderate. As a bonus, this truncation reduces
drastically the spurious center of mass components of our wave
functions that can therefore be controlled perturbatively. Hence,
our valence space will encompass the 1s$_{1/2}$, 0d$_{3/2}$, 0f$_{7/2}$,
1p$_{3/2}$, 0f$_{5/2}$ and 1p$_{1/2}$ orbits, leading to maximum 
basis sizes of  O(10$^{9}$). Following the notation of ref.~\cite{revmoph}
we call this valence space $r_2pf$. In an harmonic oscillator major shell of principal
quantum number $p$, $r_p$ represents all the orbits except the one with
the largest $j=p+\frac{1}{2}$.

For this valence space we will use the effective interaction of
ref.\cite{Ar36exp}, named SDPF.SM and described in detail in \cite{revmoph}.
In addition, to reduce the effects of the mixing with states with spurious
centre of mass excitations, we add to the interaction 
the center of mass hamiltonian, $\lambda_{CM}H_{CM}$, with $\lambda_{CM}=0.5$.
The effect of this correction is small, since the blocking of the
$d_{5/2}$ orbit reduces greatly the centre of mass contamination.
An unwelcome effect of the blocking of the  $d_{5/2}$ orbit, we have to live with,
is that it does not affect equally  to the different np-nh configurations.
It has none  on  the closed shell and becomes maximal for the
12p-12h excitations. In ref.\cite{Ar36exp} the losses of
correlation energy were compensated by an attractive 
monopole term parabolic in $n$. We shall do the
same in the present case, but, instead of trying to infer from
partial calculations the parameters of the monopole correction,
we shall adjust them so as to locate the two excited 0$^+$ states
close to their experimental values. We discuss this issue in more detail
along with the presentation of the results of the fully 
mixed  calculations.

\section{Calculations at fixed n{\hw}}
\label{sec:fix}

\begin{figure}
\begin{center}
\includegraphics[%
  width=0.7\columnwidth,
  angle=270]{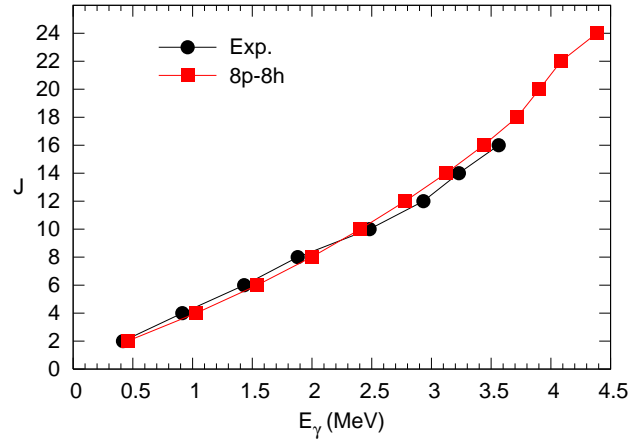}\end{center}
\caption{\label{cap:fig_sd8ph}Superdeformed band in $^{40}$Ca. $E_{\gamma}$'s,
experimental vs fixed 8{\hw} results}
\end{figure}

With the effective interaction and the valence space ready, we start
by making calculations at fixed n{\hw}. The aim is to verify that our expectations based 
in previous experiences and in algebraic models  are fulfilled.
And indeed they are.  In fig.\ref{cap:fig_sd8ph} we present the results of 
the 8{\hw} calculation compared with the experimental superdeformed
band from ref.\cite{Ca40exp} in a backbending plot.
The accord is excellent, only
the slight change of slope at $J=10$  is not reproduced by the calculation.
Notice that the band is very regular, showing no backbending up to
$J=24$, contrary to the situation in $^{48}$Cr or $^{36}$Ar. This
delay in the alignement regime is surely due to the extra collectivity
induced by the presence of  eight particles in quasi-$SU(3)$ orbitals
-four in $^{36}$Ar- and four particles in pseudo-$SU(3)$ orbitals-
absent in $^{48}$Cr. The tiny backbending at J=20 produced by the
truncated calculation of ref. \cite{Ca40nph}, disappears when the
unrestricted $r_2pf$ valence space is employed.

In table \ref{cap:tab_sd8ph} we have collected the quadrupole properties
of the superdeformed band. We have computed the $B(E2)$'s and the
spectroscopic quadrupole moments $Q_{spec}$ using the conventional
effective charges $\delta q_{\pi}=\delta q_{\nu}=0.5$ and b=1.974~fm.
We extract the intrinsic (transition) quadrupole moments $Q_{0}(t)$ 
from  the $B(E2)$'s and the static ones $Q_{0}(s)$
from the spectroscopic quadrupole moments, assuming $K=0$ and using
the standard formulas relating laboratory and intrinsic frame quantities;

\begin{equation}
\label{bmq}
 Q_0(s)=\frac{(J+1)\,(2J+3)}{3K^2-J(J+1)}\,Q_{spec}(J), \quad K\ne 1\\
\end{equation}
\begin{eqnarray}
 B(E2,J\rightarrow J-2)=
\frac{5\,e^2}{16\pi}|\langle JK20|
 J-2,K\rangle |^2 \, Q_0^2(t) 
\label{bme2}
\end{eqnarray}
$$  \quad K\ne 1/2,\, 1; $$

\begin{figure}
\begin{center}\includegraphics[%
  width=0.7\columnwidth,
  angle=-90]{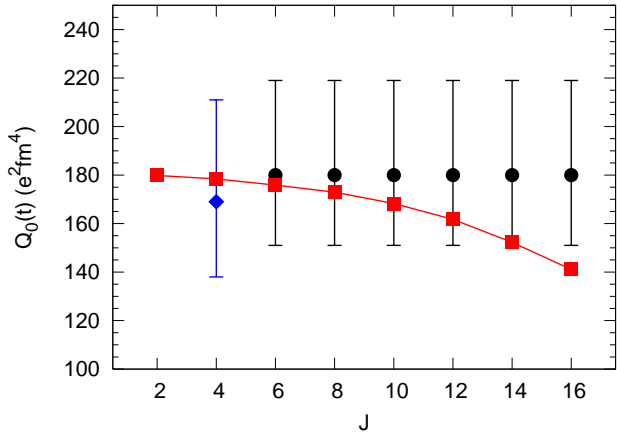}\end{center}
\caption{\label{cap:fig_Q8ph}Superdeformed band in $^{40}$Ca; transition quadrupole
  moments Q$_{0}$(t);  experimental results from ref.~\cite{woods} (lozenges) and
  from ref.\cite{Ca40exp} (circles) vs the results of the fixed 8p-8h calculation (squares).}
\end{figure}

The value of the intrinsic quadrupole moment roughly corresponds to
a deformation parameter $\beta=0.6$, which is characteristic of a superdeformed
shape and agrees with the experimental value of ref.\cite{Ca40exp}
(a comparison with a subsequent analysis of the same experiment \cite{Ca40exp2}
will be made in sections \ref{sec:sdband} and \ref{sec:outband}).
In Fig.~\ref{cap:fig_Q8ph} we have plotted
the calculated and experimental results; they agree within the large experimental
error bars. Notice that as $J$ grows, the theoretical
results lose some collectivity, whereas the experimental fit to the 
Doppler Shift Attenuation data is compatible with a constant transition quadrupole 
moment. The experimental point corresponding to the 4$^+$$\rightarrow$2$^+$ 
transition comes from an earlier measure of the lifetime of the  4$^+$ state and 
its in-band branching ratio \cite{woods}.

\begin{table}
\begin{singlespace}
\caption{\label{cap:tab_sd8ph}Quadrupole properties (in efm$^{2}$ and e$^{2}$fm$^{4}$)
of the superdeformed band in $^{40}$Ca in a fixed 8p-8h
calculation.}
\end{singlespace}
\begin{singlespace}
\begin{center}\begin{tabular}{ccccc}
\hline 
J&
Q$_{spec.}$&
B(E2)$_{J\rightarrow J-2}$&
Q$_{0}$(s)&
Q$_{0}$(t)\tabularnewline
\hline
2&
-51.4&
643 &
180&
180\tabularnewline
4&
-64.6&
905&
178&
178\tabularnewline
6&
-68.4&
968&
171&
176\tabularnewline
8&
-69.6&
980&
165&
173\tabularnewline
10&
-69.9&
953&
161&
168\tabularnewline
12&
-70.5&
896&
159&
162\tabularnewline
14&
-72.0&
804&
159&
152\tabularnewline
16&
-72.5&
696&
159 &
141\tabularnewline
\hline
\end{tabular}\end{center}\end{singlespace}

\end{table}

 The rather impressive agreement of the energetics of the superdeformed
band with the calculation at fixed 8{\hw} is reminiscent of what we had
found in the study of  $^{36}$Ar and, as it was the case then, suggests that the
SD band is of rather pure 8p-8h character.

\begin{figure}
\begin{center}\includegraphics[%
  width=0.7\columnwidth,
  angle=270]{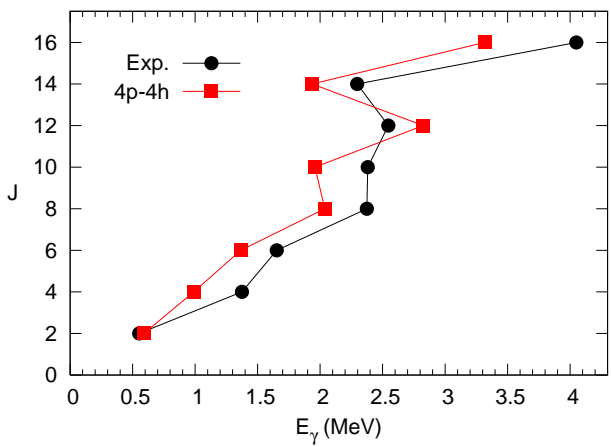}\end{center}
\caption{\label{cap:fig_nd4ph}ND band in $^{40}$Ca. $E_{\gamma}$'s,
experiment vs 4p-4h calculation.}
\end{figure}

We have  performed  as well a fixed 4p-4h calculation in order to describe the
ND  bands in $^{40}$Ca.
The most salient aspect of the calculated results is the triaxial character of the
solution, with a well developped $\gamma$-band based in the second
$2^{+}$ state ($2_{\gamma}^{+}$) of the 4p-4h configuration.
The results for the band based on the lowest 0$^+$  state of the 4{\hw}
space are compared
to the experimental ones from ref.~\cite{Ca40exp} (band 2)
in Fig.~\ref{cap:fig_nd4ph}. The agreement is now much worse that
for the SD band, although the main trends are already present in the
calculation. 
At  $J=12$ the experimental band upbends while
the calculation  produces a strong backbending.

\begin{table}

\caption{\label{cap:tab_gam4ph}Comparison of the experimental transition energies (in
keV) in the $\gamma$ band of $^{40}$Ca with the fixed 4{\hw} calculation.}

\begin{center}\begin{tabular}{ccc}
\hline 
Transition&
E$_{\gamma}$(4p-4h)&
E$_{\gamma}$(Experiment)\tabularnewline
\hline
$3_{\gamma}^{+}\rightarrow2_{\gamma}^{+}$&
819&
781\tabularnewline
$4_{\gamma}^{+}\rightarrow2_{\gamma}^{+}$&
1244&
1260\tabularnewline
$5_{\gamma}^{+}\rightarrow3_{\gamma}^{+}$&
1187&
1369\tabularnewline
$7_{\gamma}^{+}\rightarrow5_{\gamma}^{+}$&
1501&
1538\tabularnewline
$9_{\gamma}^{+}\rightarrow7_{\gamma}^{+}$&
2346&
2773\tabularnewline
$11_{\gamma}^{+}\rightarrow9_{\gamma}^{+}$&
1518&
1827\tabularnewline
$13_{\gamma}^{+}\rightarrow11_{\gamma}^{+}$&
1943&
3044\tabularnewline
\hline
\end{tabular}\end{center}
\end{table}

 The 2$^+$ band-head of the $\gamma$-band is located 2.05~MeV above the 0$^+$
band-head of the K=0 ND band. Experimentally the splitting is 1.90~MeV. The
transition energies inside the $\gamma$-band are compared to the experimental
ones,  ref.~\cite{Ca40exp} (band 4),  in table \ref{cap:tab_gam4ph}.
 The calculated values compare reasonably well with the experiment, except
for the location of the 13$^+$ state that appears 1~MeV too low.

The quadrupole properties of the K=0 ND band
are shown in table \ref{cap:tab_nd4ph}. In this case, the
corresponding deformation parameter has the value $\beta=0.3$, a
typical value for a normal deformed band. We postpone the comparison with
the experimental data to the discussion of the complete calculation.

\begin{table}
\begin{singlespace}
\caption{\label{cap:tab_nd4ph}Quadrupole properties (in efm$^{2}$ and e$^{2}$fm$^{4}$)
of the ND of $^{40}$Ca in a fixed 4{\hw} calculation.}
\end{singlespace}
\begin{singlespace}
\begin{center}\begin{tabular}{ccccc}
\hline 
J&
Q$_{spec.}$&
B(E2)$_{J\rightarrow J-2}$&
Q$_{0}(s)$&
Q$_{0}(t)$\tabularnewline
\hline
2&
-31.0&
269&
109&
116\tabularnewline
4&
-41.0&
364&
113 &
113\tabularnewline
6&
-48.9&
341&
120&
104\tabularnewline
8&
-44.0&
309&
104&
97\tabularnewline
10&
-48.9&
237&
113&
84\tabularnewline
12&
-38.3&
86&
86&
50\tabularnewline
14&
-40.9&
115&
91&
58\tabularnewline
16&
-34.8&
47 &
76&
37\tabularnewline
\hline
\end{tabular}\end{center}\end{singlespace}

\end{table}

We close this section by stressing that
the shell model calculations in the $r_2pf$ valence space at
fixed 8p-8h and 4p-4h configurations give a convincing description of the superdeformed
band in $^{40}$Ca, and are also able to describe the corresponding
ND rotational spectra,  which turns out to be that of a triaxial structure 
developing a K=0 and a K=2 ($\gamma$) band, in agreement with the experimental information.
Indeed, in the physical states the configurations with different values of
n{\hw}  mix, mainly through the cross shell pairing interactions,
and it will be the task of the next sections to understand how this mixing
proceeds and to compare the complete results with the experimental data.

\section{The energies of the n{\hw}   band-heads relative to the closed shell}
\label{sec:poz0}
 
Before moving into the full  $r_2pf$ space diagonalizations, we need to 
explore more in depth the information gathered in the fixed N{\hw}   calculations.
With the  SDPF.SM interaction, the n{\hw}   bandheads lie too high in energy
relative to the closed shell, as it was the case in our study of $^{36}$Ar.
A small part of this missing energy can be due to residual defects of the cross-shell
monopole terms of the interaction, but the bulk of it is due to the blocking
of the 0d$_{5/2}$ orbit. We can absorb this effect via the modification of two
global monopole terms: 

\begin{equation}
\Delta(n) = \frac{1}{2} n (n-1) \delta_1 + n (12-n) \delta_2
\label{eqn:d5}
\end{equation} 

\noindent
where $n$ is the number of particles in the $pf$-shell, and the
$\delta$'s are the modifications of the global $pf$-shell monopole 
interaction and the $r_2$--$pf$
monopole interaction, respectively. In our study of $^{36}$Ar, an equivalent
expression was used. In a first set of exploratory calculations,
we realized that when the $\delta$'s were chosen so as to locate the
two excited 0$^+$ states at about their experimental excitation
energies, the percentage of closed
shell in the ground state of $^{40}$Ca was too low (54\%), {\it i. e.} the mixing
was too strong.  In order to diminish the mixing we have multiplied all the
cross shell off-diagonal matrix elements by a factor 0.8. Now, the
values of the $\delta$'s that place the  0$^+$ states correctly 
are $\delta_1$=--0.27~MeV,
and $\delta_2$=--0.130~MeV. With this choice the
closed shell component of the ground state rises to 65\%, that we take
as a reasonable value. We have used another  set of $\delta$'s
that gives 75\% of closed shell and no dramatic changes are observed
in the ensemble of  observables.  As the 0.8 scaling implies a reduction of
the pairing interactions in the space, the isovector pairing matrix
elements of the $pf$-shell part of the SDPF.SM interaction, that were
reduced by a factor 0.6 in our study of $^{36}$Ar, are taken now as
0.8 times their SDPF.SM values. In summary,
we have loosely fitted two global monopoles of the 
interaction to the location of the excited
0$^+$ states, keeping the core excited components in the 30\% range.
Actually, the difference in correlation energy between a full $sd$-shell
calculation for the nucleus $N=Z=20-\frac{n}{2}$, and the calculation with the orbit 0d5/2
blocked, accounts for two thirds of the value of $\Delta(n)$. 
The pairing modifications can be justified with perturbation
theory arguments (see the discussion in \cite{Ar36}).

\begin{figure}
\begin{center}\includegraphics[%
  width=1.0\columnwidth,
  angle=0]{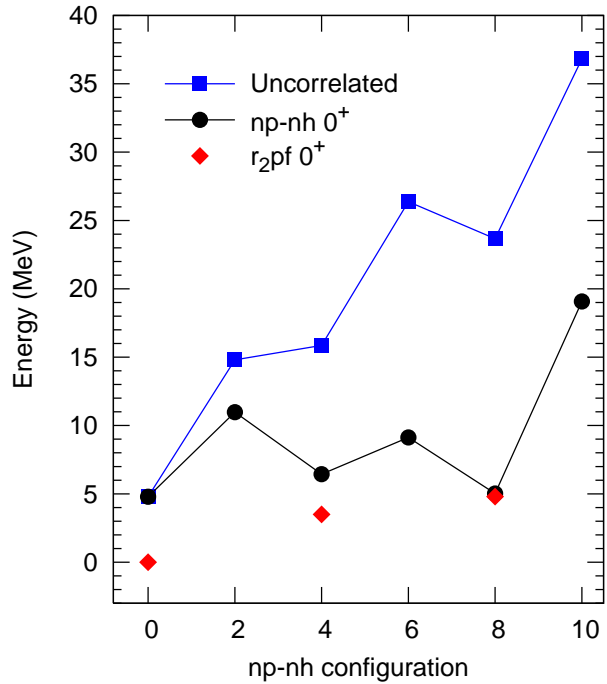}\end{center}
\caption{\label{cap:fig_pot0} Energies in the different np-nh configurations;
Lowest Slater Determinant plus the
correction of equation \ref{eqn:d5} (squares); Energy of the lowest 0$^+$ state
in the $r_2pf$ space at fixed N{\hw} (circles). The diamonds give the
energies of the fully mixed  calculations in the  $r_2pf$ space}
\end{figure}

We have now all the ingredients needed to make quantitative the discussion of sec.~\ref{sec:sph}.
They are plotted in Fig.~\ref{cap:fig_pot0}. 
The squares
give the lowest uncorrelated energy in each N{\hw}  space (to be more precise, the minimum of the expectation value
of the SDPF.SM Hamiltonian plus the correction of equation~\ref{eqn:d5}, 
calculated for all  the  Slater Determinants of the m-scheme spherical basis).
 As we had anticipated the uncorrelated
energies do not grow linearly with the number of particles excited across the N=Z=20 gap
(if it had been so, all the squares would lie in the prolongation of the straight line 
joining the n=0 and n=2 points). The
increase is much slower, with a superimposed odd-even effect in n/2. Notice that energy of the lowest
8{\hw}  Slater Determinant lies $\approx$~20~MeV above the closed shell.   
Finally, we carry out the 
explicit diagonalizations separately for each N{\hw}  value and we find the results plotted as circles.
We can see that the correlation energies are very large, in particular for the SD band-head
that gains 18.5~MeV and becomes almost degenerated with the closed shell configuration.
We can make a rough analysis of the correlation energies in terms of their T=0 and T=1 pairing and
multipole-multipole (mainly quadrupole-quadrupole) content. For that we compute the expectation value
of the (monopole free) pairing part of the effective interaction in the fixed n{\hw}  bandheads and 
subtract it from the total correlation energy. In the 4{\hw} 0$^+$ ND state, out of the 9.5~MeV of
correlation energy, 5.5~MeV come from T=1 pairing, 0.5~MeV from T=0 pairing and 3.5~MeV
from the quadrupole correlations. In the 8{\hw} 0$^+$ SD state, the share is 5.5~MeV,
0.5~MeV and 12.5~MeV respectively. Thus, the contribution of the neutron proton
pairing amounts to 2.33~MeV for any of the bands.

The structure of the line joining the black circles in the plot in
quite interesting, because it resembles the energy versus deformation
curves typical of projected Hartree-Fock calculations. The number of
particle-hole excitations in the X-axis can be taken as roughly
proportional to the deformation. With this perspective, we
distinguish three coexisting minima, spherical, n=0, deformed, n=4, and
superdeformed, n=8, separated by the n=2 and n=6 maxima.  The n=10
band-head lies very high in energy and we have verified that it plays
no role at all in this problem. The structure of this ``potential
energy'' curve, consisting of wells and barriers, provides an
intuitive picture of how the deformed and superdeformed bands can
preserve their identity after mixing.

\begin{figure}
\begin{center}\includegraphics[%
  width=1.0\columnwidth,
  angle=0]{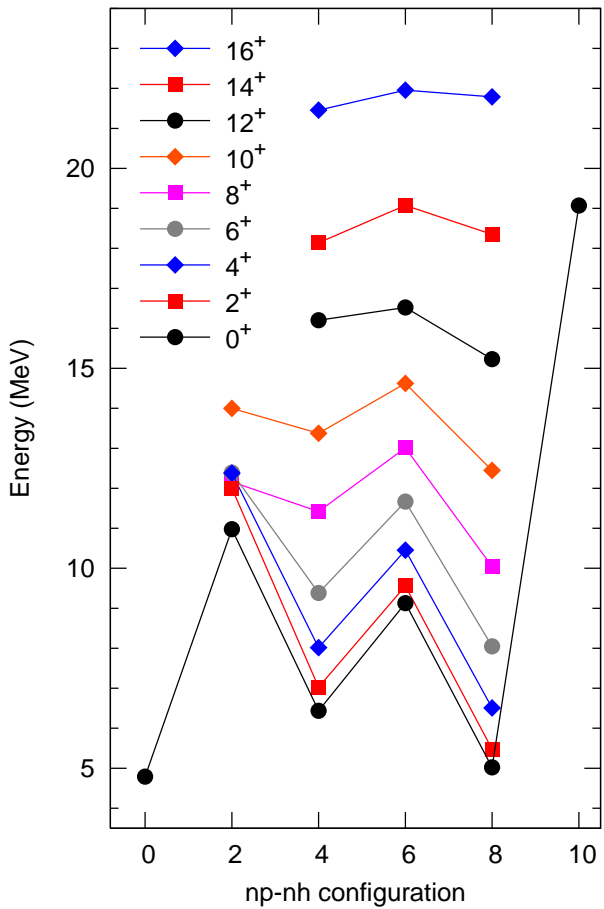}\end{center}
\caption{\label{cap:fig_potJ} Energies of the yrast states in the fixed np-nh calculations
(corresponding to the circles in Fig.~\ref{cap:fig_pot0} for J=0)}
\end{figure}

Keeping on with the mean field analogy, it is worth to note that the black circles in  Fig.~\ref{cap:fig_pot0}
correspond to J=0$^+$ states, {\ i. e.} they are angular momentum projected solutions  
(before variation, VAP in the usual jargon).
In  Fig.~\ref{cap:fig_potJ} we add the points for all the lowest states of even J  that appear in the deformed
(4{\hw}  and 6{\hw})  and superdeformed (8{\hw}) bands and the corresponding 2{\hw}  and 10{\hw}  states.
We can observe that the structure of wells and barriers of the J=0 curve, that protects the
deformed and superdeformed bands from strong mixing,
is maintained up to J=10. Above, the barriers flatten and we should expect larger mixing.
In particular this can lead to the erosion of the collectivity of the SD band at high spin
in the calculation.

\section{Results of the complete calculations in the $r_2pf$ space}
\label{sec:full}

 The calculations in the  $r_2pf$ space are computationally very demanding. In addition to
the very large dimensions involved, the need to compute several states of the same 
angular momentum increases substantially the number of Lanczos iterations needed to
achieve convergence. The shell model code {\sf ANTOINE} \cite{revmoph} has been used throughout.

 The structure of the   first three $0^{+}$ states, the spherical ground state and 
the excited deformed and super-deformed band-heads is shown in table \ref{tab:npnh1}.
The two body cross-shell off diagonal matrix elements can connect directly configurations
differing only in two particle hole jumps. We see in the table that, indeed, the ground
state 0p-0h leading component is mainly correlated by 2p-2h components. The leading
4p-4h component of the ND state can, in principle mix with 2p-2h  and 
6p-6h  components. Actually the mixing is dominated by the latter, with a non-negligible
8p-8h piece. The leading 8p-8h component of the SD could also mix with 6p-6h and
10p-10h configurations directly, but it chooses none. The SD band-head is very pure,
with only small amounts of 4p-4h and 6p-6h components. Let us mention  that
the yrast band in the 6{\hw}  space, corresponds also to a rotor with deformation slightly larger
than the one in the 4{\hw} space. On the contrary, the deformation and the correlation energy   
in the 10{\hw} space are smaller, leading to negligible mixing with the other spaces.

\begin{table}

\caption{Percentage of np-nh components and energy of the first three $0^{+}$states (GS, ND, and SD) of $^{40}$Ca}

\begin{center}\begin{tabular}{cccccccccc}
\hline 
&
0p-0h&
2p-2h&
4p-4h&
6p-6h&
8p-8h&
&
&
E(th)&
E(exp)\tabularnewline
\hline
$0_{GS}^{+}$&
65&
29&
5&
-&
-&
&
&
0&
0\tabularnewline
$0_{ND}^{+}$&
1&
1&
64&
25&
9&
&
&
3.49&
3.35\tabularnewline
$0_{SD}^{+}$&
-&
-&
9&
4&
87&
&
&
4.80&
5.21\tabularnewline
\hline
\end{tabular}\end{center}
\label{tab:npnh1}
\end{table}

 In the process of mixing the winner is, energy wise, the ground state that
gains almost 5~MeV, mostly pairing-like. The ND band gains barely 2~MeV
and the SD band essentially nothing. That's why, in order to reproduce
the experimental situation, the three 0$^+$ states before
mixing must be degenerated or even with their energies inverted 
The energies of the three physical  0$^+$ states after mixing are
represented by the diamonds in Fig.~\ref{cap:fig_pot0}.

\subsection{The Super-Deformed Band}
\label{sec:sdband}

\begin{figure}
\begin{center}\includegraphics[%
  width=0.7\columnwidth,
  angle=270]{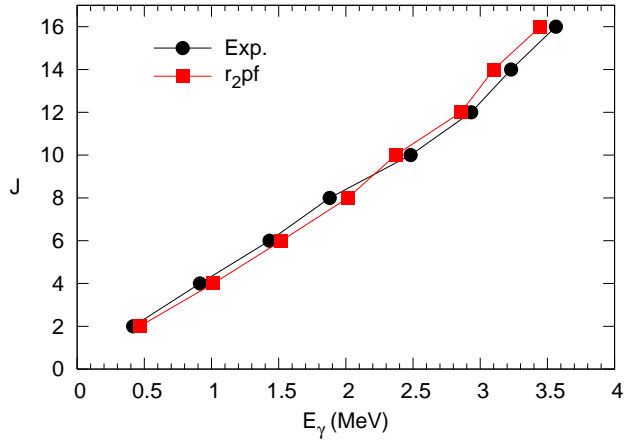}\end{center}
\caption{$E_{\gamma}$'s in the SD band, mixed calculation $vs.$ experiment\label{cap:fig_sdband}}
\end{figure}

 We start the discussion of the results with the excitation energies of the
SD band as produced by the fully mixed calculation. As we can see in 
Fig.\ref{cap:fig_sdband}, the mixing does not modify noticeably the 
features already present in the fixed 8p-8h calculation. Perhaps one could detect
some irregularities at the upper part of the band. The calculated sequence
crosses the experimental one at around the rotational frequency where the
calculated states start loosing collectivity, but in global terms, the 
agreement is excellent. The results of the mixed calculation beyond J=16
are equivalent to that of the fixed 8{\hw} calculation shown in Fig. \ref{cap:fig_sd8ph}

 In table~\ref{tab:npnh2}, we give the np-nh structure of the SD band up to J=16.
The 8p-8h component is nicely constant up to J=10, as we expected from the analysis
of the potential energy curves in Fig.\ref{cap:fig_potJ}. At higher angular momentum,
the mixing with 6{\hw} components becomes stronger, and we should expect a less collective
behavior. It is interesting to follow the evolution of the location of the SD states
with increasing angular momentum. The J=0 SD state is the third J=0 state in the 
spectrum. The J=2, 4, and 6 states are the second of their spins, J=8 and 10, third,
J=12, fourth, J=14 sixth, and finally J=16 fourth again.  

\begin{table}

\caption{np-nh structure of the superdeformed band of $^{40}$Ca.}
\label{tab:npnh2}
\begin{center}\begin{tabular}{cccccc}
\hline 
J&
0p-0h&
2p-2h&
4p-4h&
6p-6h&
8p-8h\tabularnewline
\hline
0&
-&
-&
9&
4&
87\tabularnewline
2&
-&
-&
11&
4&
85\tabularnewline
4&
-&
-&
8&
5&
87\tabularnewline
6&
-&
-&
3&
5&
91\tabularnewline
8&
-&
-&
2&
6&
91\tabularnewline
10&
-&
-&
1&
12&
87\tabularnewline
12&
-&
-&
2&
29&
69\tabularnewline
14&
-&
-&
11&
27&
63\tabularnewline
16&
-&
-&
0&
40&
60\tabularnewline
\hline
\end{tabular}\end{center}
\end{table}

We shall examine now the quadrupole properties of the
superdeformed band in $^{40}$Ca as they come out of the  full $r_2pf$ calculation.
In table \ref{tab:be2sd} we have gathered the theoretical spectroscopic quadrupole 
moments and the B(E2)
values. We can conclude that the mixing  causes just an erosion of the 
8{\hw} values presented in table \ref{cap:tab_sd8ph} up to J=10-12. Beyond, the larger presence
of less collective 6{\hw} components has a much stronger effect, particularly
in the B(E2)'s. Thus, the values of the transition quadrupole moments diminish
rapidly at the end of the band. The static quadrupole moments vary less abruptly.

\begin{table}
\begin{singlespace}
\caption{\label{tab:be2sd}Quadrupole properties (in efm$^{2}$ and e$^{2}$fm$^{4}$) of the
superdeformed band in $^{40}$Ca. Full $r_2pf$ calculation.}
\end{singlespace}
\begin{singlespace}
\begin{center}\begin{tabular}{ccccc}
\hline 
J&
Q$_{spec.}$&
B(E2)$_{J\rightarrow J-2}$&
Q$_{0}$(s)&
Q$_{0}$(t)\tabularnewline
\hline
2&
 -47.7&
579&
167 &
171\tabularnewline
4&
-61.1&
813&
168&
169\tabularnewline
6&
-66.3 &
874&
166&
167\tabularnewline
8&
-66.5&
906&
158&
166\tabularnewline
10&
-66.3&
844&
153&
158\tabularnewline
12&
-71.8&
546&
162&
126\tabularnewline
14&
-62.1&
557&
139&
127
\tabularnewline
16&
-64.9&
429&
142 &
111\tabularnewline
\hline
\end{tabular}\end{center}\end{singlespace}

\end{table}

\begin{figure}
\begin{center}\includegraphics[%
  width=0.7\columnwidth,
  angle=270]{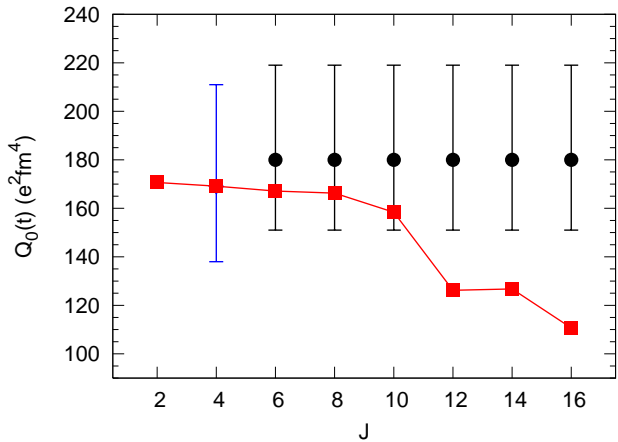}\end{center}
\caption{\label{cap:fig_q0sd}Transition quadrupole moments in the SD band. Full $r_2pf$
results (squares) compared to the experimental data from refs. \cite{woods} (lozenges) and
\cite{Ca40exp} (circles)}
\end{figure}

In Fig.~\ref{cap:fig_q0sd} we compare our calculated numbers with the experimental
results. Comparing with the 
fixed 8{\hw} results, we find a large reduction of the transition quadrupole moments
of the three upper transitions of the band. It is possible that this drop in
collectivity be an  artifact of the $r_2pf$ space, but for the moment we have no way
to verify it.

 The experimental points in the figure for J$\ge$6 are those of the first analysis of the
Doppler Shift Attenuation data made in ref.\cite{Ca40exp}. The J=4 point comes from another, earlier,
experiment \cite{woods}. The calculated transition quadrupole moments agree with the data
 for J$\le$10, but underestimate them for the three uppermost transitions. In
ref.\cite{Ca40exp2} a reanalysis of the same experimental data was made, and the preferred
solution differs from the precedent one. Whereas the former solution was a constant
Q$_0$(t)=180$^{+0.39}_{-0.29}$~e~fm$^{2}$ for the six transitions from the states of the
SD band with J=16 to J=6, the latter gives Q$_0$(t)=181$^{+0.41}_{-0.26}$$\pm0.21$~e~fm$^{2}$ for the
J=16 to J=12 states and Q$_0$(t)=118$^{+0.06}_{-0.05}$$\pm0.13$~e~fm$^{2}$
for J=10 to J=6. Notice that, compared to ref.\cite{Ca40exp} an extra systematic error has been added.
In spite of that, the final error bars of the values corresponding to the
lower J transitions are largely reduced with respect to those of the upper transitions.
Actually, their extracted Q$_0$(t) values are barely compatible with the value for the 4$^+$$\rightarrow$2$^+$  914~keV transition,
Q$_0$(t)=169$^{+0.42}_{-0.32}$~e~fm$^{2}$, obtained in ref. \cite{woods} (the lower and upper tips of one and
another error bars just touch at the value 137~e~fm$^{2}$).

 If the results of the analysis of Chiara {\it et. al.} were the only viable interpretation
of the data, the comparison with our calculated
 Q$_0$(t)'s. would be rather poor, meaning that something important in missing in our approach. 
The beautiful agreement we had for J$\le$10 will be lost, and the fact that the increase of
the experimental error bars for the upper transitions make our results deviate less, is a meager
compensation. We have tested solutions with larger mixing, through the mechanisms 
discussed in section \ref{sec:poz0}, but even going as far as
keeping just 50\% of closed shell in the ground state, we obtain only an extra 10\% erosion of the
B(E2)'s for J$\le$10. On the contrary, more mixing brings in large reductions of the transitions in the upper
part of the SD band. In the section dealing with the out-band decay branches we propose a mechanism
that conciliates the theoretical picture with the experimental data. It is based on a detailed
analysis of the decay of the J=8 member of the SD band, a decay that  
in our opinion, has driven the fit of  ref.\cite{Ca40exp2} into the new set of  Q$_0$(t)'s.

\subsection{The Normal-Deformed Triaxial Bands}
\label{sec:ndband}

 We move now to the full $r_2pf$ space results for the two bands that according to our calculations
have a dominant 4p-4h structure. As we have shown in section \ref{sec:fix} they are consistent
with the presence of a deformed triaxial intrinsic state. In table \ref{tab:4p4h} we list
the percentage of the different np-nh components in the
deformed band of $^{40}$Ca based on the first excited 0$^+$ state.
 We observe that the 4p-4h dominance is less strong than the 8p-8h dominance in the
SD band at low spins, and larger at high spins. Another characteristic feature is that the
mixing proceeds through the 6p-6h components, with the 2p-2h components completely absent except
in the J=8 state, where an accidental degeneracy occurs.
 We do not have a definite explanation
for this preference, that can be due to phase space considerations (there are much more 6p-6h
states to mix with than 2p-2h states) but more probably to the fact that the collectivity of the 4p-4h 
and 6p-6h spaces is very similar and much larger than that of the 2p-2h space.
That the mixing strength of the
6p-6h space is exhausted by the ND band  could explain why the SD band is so pure.  
As expected in a collective picture, the spread of the wave functions of the states of the 
$\gamma$ band among the np-nh spaces is very similar to that of the ND band.

\begin{table}
\caption{\label{tab:4p4h}np-nh structure of the deformed band of $^{40}$Ca based on the first excited 0$^+$ state}
\begin{center}\begin{tabular}{cccccc}
\hline 
J&
0p-0h&
2p-2h&
4p-4h&
6p-6h&
8p-8h\tabularnewline
\hline
0&
1&
1&
64&
25&
9\tabularnewline
2&
-&
1&
64&
24&
10\tabularnewline
4&
-&
1&
68&
23&
8\tabularnewline
6&
-&
2&
75&
20&
3\tabularnewline
8&
-&
21&
62&
15&
2\tabularnewline
10&
-&
-&
81&
17&
1\tabularnewline
12&
-&
-&
81&
18&
1\tabularnewline
14&
-&
-&
82&
17&
1\tabularnewline
16&
-&
-&
79&
19&
1\tabularnewline
\hline
\end{tabular}\end{center}
\end{table}

\begin{figure}
\begin{center}\includegraphics[%
  width=0.7\columnwidth,
  angle=270]{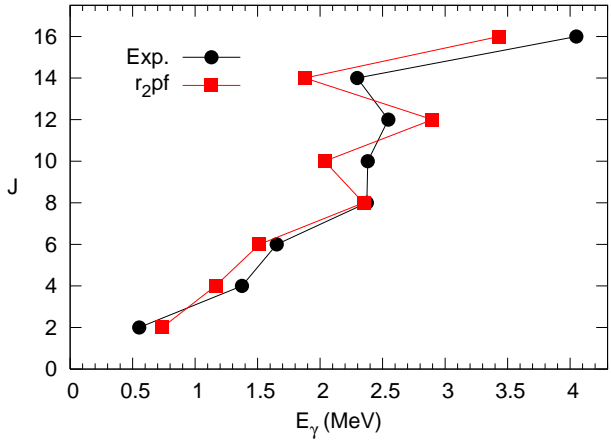}\end{center}
\caption{\label{cap:fig_ndband}Deformed band in $^{40}$Ca. $E_{\gamma}$'s,
experiment $vs$ full $r_2pf$ calculation.}
\end{figure}

 In Fig.~\ref{cap:fig_ndband} we plot the calculated energies of the ND band 
compared with the experiment. We observe that the mixing
improves clearly the agreement  in  the lower part of the band. The discrepancy at 
the backbending is the same already present in the 4{\hw} calculation.

 The 2$^+$ band-head of the $\gamma$ band is predicted at 5.88~MeV of excitation energy
compared with the experimental value 5.25~MeV. This means that the mixing  increases
the splitting of the ND and $\gamma$ bands by 350~keV with respect to the
result of the 4{\hw} calculation. The in-band excitation energies change
very little with respect to the 4{\hw} values gathered in table \ref{cap:tab_gam4ph}.

\begin{table}
\caption{\label{tab:ndbe2}Quadrupole properties (in efm$^2$ and e$^2$fm$^4$) of the ND
 $^{40}$Ca, $r_2pf$ calculation}
\begin{center}\begin{tabular}{ccccc} 
\hline  
J& Q$_{spec.}$& B(E2)$_{J\rightarrow J-2}$& Q$_{0}$(s)& Q$_{0}$(t)\tabularnewline 
\hline   
2& -32.2& 292& 113& 121\\  

4& -42.1& 397& 116& 118\\  

6& -47.3& 346& 118& 105\\  

8& -35.5& 227&  84&  83\\ 

10& -48.1& 161& 111&  69\\ 

12& -37.6&  75&  85&  47\\ 

14& -39.1& 112&  87&  57\\ 

16& -35.6&  49&  78&  37\\ 

\hline \end{tabular}\end{center}
\end{table}

\begin{figure}
\begin{center}\includegraphics[%
  width=0.7\columnwidth,
  angle=270]{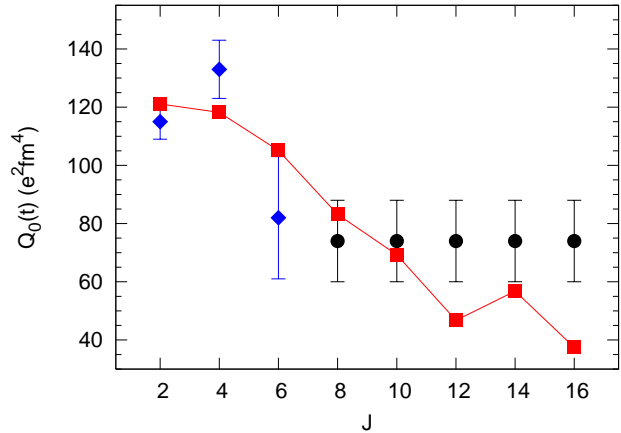}\end{center}
\caption{\label{cap:fig_q0_nd}Transition quadrupole moments in the SD band. Full $r_2pf$
results (squares) compared to the experimental data from refs. \cite{toi} (lozenges) and
\cite{Ca40exp} (circles)}
\end{figure}

 The quadrupole properties of the ND band --table \ref{tab:ndbe2}-- 
are very similar to those of the 4{\hw} calculation.
Only at J=8 there is a reduction in collectivity due to the
accidental mixing with a nearby 2p-2h state that we have already
commented. As can be seen in Fig.~\ref{cap:fig_q0_nd}, where
we have plotted the transition quadrupole moments, 
this reduction goes in the direction demanded by the
data.
Globally the agreement is quite good, the trends are very well 
reproduced and in most cases the theoretical numbers fall inside
the experimental error bars. At the upper part of the band
the theoretical values underestimate the experimental ones. 
The data come from different sources, for the J$\ge$8 they
are taken from the  Doppler Shift Attenuation analysis of ref.~\cite{Ca40exp},
for J=6 and J=8 from the lifetimes and branching ratios
 measured in refs.\cite{woods,toi}.  

In table \ref{tab:be2gam} we have collected the calculated quadrupole properties
of the $\gamma$ band. The intrinsic information has been extracted from the B(E2)'s
and spectroscopic quadrupole moments assuming K=2.
We find again a well defined deformed intrinsic state, with similar static and transition quadrupole
moments, both very similar to those of the ND band.

\begin{table}
\caption{\label{tab:be2gam}Quadrupole properties (in efm$^{2}$ and e$^{2}$fm$^{4}$) of the
$\gamma$ band in $^{40}$Ca, $r_2pf$ calculation.}

\begin{center}\begin{tabular}{cccccc}
\hline 
J&
Q$_{spec.}$&
B(E2)$_{J\rightarrow J-2}$&
B(E2)$_{J\rightarrow J-1}$&
Q$_{0}$(s)&
Q$_{0}$(t)\tabularnewline
\hline
2&
28.6&
&
&
100&
\tabularnewline
3&
-0.24&
&
427&
-&
\tabularnewline
4&
-17.7&
133&
284&
122&
106\tabularnewline
5&
-25.7&
214&
&
111&
106\tabularnewline
7&
-33.5&
211&
&
104&
90\tabularnewline
9&
-35.6&
175&
&
96&
77\tabularnewline
11&
-45.9&
110&
&
115&
59\tabularnewline
13&
-40.5&
80&
&
97&
49\tabularnewline
\hline
\end{tabular}\end{center}
\end{table}

\subsection{Out-band transitions}
\label{sec:outband}

The experimental data on out-band transitions are scarce and very often affected of
large uncertainties. For the low energy part of the spectrum of $^{40}$Ca lifetimes
and branching ratios are known for some levels. These will be compared with our
predictions in tables \ref{tab:out1} and \ref{tab:out3}. For J$\ge$6, we rely in
the semi-quantitative informations of refs. \cite{Ca40exp} and \cite{Ca40exp2}. In table
\ref{tab:out1}, we focus in the out-band transitions from ND and SD
states. The balance is uneven. The transition probabilities of the decays 
from the 2$^+$ states in the ND and SD
bands to the ground state are largely under-predicted by the calculation. When we
increase the mixing they increase at most by a factor two, very far from
what would square with the data. For these small B(E2)'s it is possibly not
sensible to reason in terms of factors, but instead think that some additive contribution
is lacking in our space. It is worth to recall here that the span of B(E2)
values that we are trying to explain simultaneously is of three orders of magnitude.
The decay of the superdeformed 0$^+$ is well reproduced, and the same applies to the
decay of the  superdeformed  4$^+$ to the deformed  2$^+$. The in-band transition 
is simultaneously well accounted for (see Fig.~\ref{cap:fig_q0sd}), but
apparently we miss  badly the two remaining transitions known experimentally.
We cannot but recognize our limitations.

\begin{table}
\begin{singlespace}
\caption{\label{tab:out1}Out-band transitions from the superdeformed (SD) and normal deformed
(ND) bands of $^{40}$Ca. The energies are in keV and the B(E2)s in
e$^{2}$fm$^{4}$.}
\end{singlespace}
\begin{singlespace}
\begin{center}\begin{tabular}{cccccc}
\hline 
Transition&
&
\multicolumn{2}{c}{E$_{\gamma}$}&
\multicolumn{2}{c}{B(E2)}\tabularnewline
&
&
Theory&
Experiment&
Theory&
Experiment\tabularnewline
\hline
$2_{ND}^{+}\rightarrow$&
$0_{GS}^{+}$&
4232&
3904&
1.8&
18$\pm$1\tabularnewline
$0_{SD}^{+}\rightarrow$&
$2_{ND}^{+}$&
565&
1307&
58&
134.8$\pm$24.5\tabularnewline
$2_{SD}^{+}\rightarrow$&
$0_{GS}^{+}$&
5263&
5629&
0.1&
1.7$\pm$0.4\tabularnewline
&
$0_{ND}^{+}$&
1769&
2277&
3&
20.9$\pm$5.0\tabularnewline
$4_{SD}^{+}\rightarrow$&
$2_{ND}^{+}$&
2045&
2638&
19.4&
21$\pm$4\tabularnewline
&
$4_{ND}^{+}$&
882&
1264&
6.8&
116$\pm$34\tabularnewline
&
$2_{3}^{+}$&
397&
1294&
2.7&
176$\pm$41\tabularnewline
\hline
\end{tabular}\end{center}\end{singlespace}

\end{table}

  In table \ref{tab:out2} we present the calculated B(E2)'s, E$_{\gamma}$'s and
 the in-band  branching ratios. For the experimental information we 
draw from the article by Chiara {\it et. al.} \cite{Ca40exp2}. The calculated
branching ratios for the three uppermost transitions of the SD band
are close to 100\% in agreement with the experimental observation.
For the next transition, the predicted 96\% branching ratio looks too
large when inspecting the figure in ref.  \cite{Ca40exp2}, but the paper
does not give a figure for it. However, it is in the decay of the 8$^+$ member
of the superdeformed bands that we depart dramatically from the
experimental branching ratio, 96\% calculated $vs$ 20\% experimental.
We shall devote some space to this comparison, mainly because it is
in our opinion due to this very number that the authors of  ref.  \cite{Ca40exp2}
obtain a fit to their Doppler Shift Attenuation data in which, for J$\le$10,
the transition quadrupole moments
of the  band have values that correspond actually to a normally deformed band.
Before that, let's mention that the branching ratios of the  6$^+$ SD state seem also consistent with the
data of Chiara {\it et al.} and with a rather pure superdeformed character.
We have already seen that the calculation also describes correctly the 
 4$^+$ SD branching ratios.

\begin{table}
\caption{\label{tab:out2}Comparison between the theoretical 
in-band $(SD\rightarrow SD)$ and
out-band transition probabilities $(SD\rightarrow ND)$ 
for the states of the superdeformed
band of $^{40}$Ca with J$\ge$6.
 The energies are in keV and the B(E2)s in e$^{2}$fm$^{4}$.}
\begin{center}\begin{tabular}{cccccc}
\hline 
\multicolumn{2}{c}{Transition}&
B(E2)$_{J\rightarrow J-2}$&
E$_{\gamma}$&
E$_{\gamma}$(th)&
BR \% \tabularnewline
\hline
$6_{SD}^{+}\rightarrow$&
$4_{SD}^{+}$&
874&
1432&
1521&
\tabularnewline
&
$4_{ND}^{+}$&
43&
2695&
2403&
46\tabularnewline
$8_{SD}^{+}\rightarrow$&
$6_{SD}^{+}$&
906&
1880&
2015&
\tabularnewline
&
$6_{ND}^{+}$&
7.8&
2921&
2904&
93\tabularnewline
$10_{SD}^{+}\rightarrow$&
$8_{SD}^{+}$&
844&
2481&
2371&
\tabularnewline
&
$8_{ND}^{+}$&
12&
3030&
2929&
96\tabularnewline
$12_{SD}^{+}\rightarrow$&
$10_{SD}^{+}$&
546&
2932&
2857&
\tabularnewline
&
$10_{ND}^{+}$&
1.5&
3.590&
3750&
100\tabularnewline
$14_{SD}^{+}\rightarrow$&
$12_{SD}^{+}$&
557&
3230&
3100&
\tabularnewline
&
$12_{ND}^{+}$&
5.0&
4264&
3952&
97\tabularnewline
$16_{SD}^{+}\rightarrow$&
$14_{SD}^{+}$&
429&
3563&
3447&
\tabularnewline
&
$14_{ND}^{+}$&
0.03&
5531&
5520&
100\tabularnewline
\hline
\end{tabular}\end{center}
\end{table}

Our argument goes as follows: Given that the phase space factors favor the
out-band transition by a factor 9, and considering that the 
B(E2)'s of the $8_{SD}^{+}\rightarrow6_{SD}^{+}$ and 
$8_{ND}^{+}\rightarrow6_{ND}^{+}$ in tables \ref{cap:tab_sd8ph} and \ref{cap:tab_nd4ph} 
are 980~e$^{2}$fm$^{4}$ and 309~e$^{2}$fm$^{4}$
respectively, it is readily seen that the experimental branching ratio
cannot be reproduced unless the 9307~keV and 9856~keV experimental states
correspond to a 50\% mixing of the pure SD and ND states. Assuming that
the J=6 states are pure ND and SD, this leads to BR=27\%, with in-band
B(E2)=~490~e$^{2}$fm$^{4}$ and out-band B(E2)=~150~e$^{2}$fm$^{4}$.
The in-band transition in the ND band should have also  B(E2)=~150~e$^{2}$fm$^{4}$.
If we further assume that the J=10, 12338~keV, state of the SD band is  pure, it will 
decay equally to the 9307~keV and 9856~keV J=8 states, which seems to be the experimental
situation. Translating this into transition quadrupole moments, we should
have Q$_0$(t)(10$^+$)=118~e~fm$^{2}$ and  Q$_0$(t)(8$^+$)=122~e~fm$^{2}$ in the
super-deformed band and Q$_0$(t)(8$^+$)=69~e~fm$^{2}$ in the normally deformed
band, in excellent agreement with
the experimental analysis of refs. \cite{Ca40exp,Ca40exp2}. As a bonus, the
low spin part of the band remains truly superdeformed.
Actually the band is superdeformed all along except for the 8$^+$ state,
even if the accidental degeneracy
of the ND and SD 8$^+$ states provokes a strong reduction of the B(E2)'s of the
transitions to and from the latter  state. 
In addition, the lifetime of the  9856~keV, J=8, state is reduced by about a factor two.

Then, why are these features absent in our calculation? It is clear that in order
to obtain a 50\% mixing of the ND and SD states, they must be degenerate before mixing,
their effective splitting being not larger than about  200~keV. To match this requirement is beyond the
accuracy of our theoretical description. Paradoxically,
the calculated excitation energies of the  9307~keV and 9856~keV J=8 states, 
9260~keV and 9820~keV look astonishingly precise. But we fail to have the first
excited 8$^+$ state, experimentally at 8103~keV  at the right energy; it its
predicted at 8900~keV. This state has a 2p-2h aligned nature and the fact
that it mixes strongly with the ND state, that we have already discussed, means
that both are degenerate before mixing at about their mean excitation energy.

In a sense, the character of the SD band is closer to what is
suggested by the calculated static quadrupole moments in table
\ref{tab:be2sd}. The only modification brought in by the 8$^+$ anomaly
would be a reduction of its  Q$_{0}$(s)  from 165~e~fm$^{2}$ to
135~e~fm$^{2}$

\begin{table}

\caption{\label{tab:out3}Out-band transitions from the $\gamma$ band of $^{40}$Ca. The energies
are in keV and the B(E2)s in e$^{2}$fm$^{4}$.}

\begin{center}\begin{tabular}{cccccc}
\hline 
\multicolumn{2}{c}{Transition}&
\multicolumn{2}{c}{B(E2)}&
\multicolumn{2}{c}{E$_{\gamma}$}\tabularnewline
\multicolumn{2}{c}{}&
Theory&
Experiment&
Theory&
Experiment\tabularnewline
\hline
$2_{\gamma}^{+}\rightarrow$&
$0_{GS}^{+}$&
0.20&
1.0$\pm$0.3&
5690&
5249\tabularnewline
&
$0_{ND}^{+}$&
16&
100$\pm$3&
2277&
1896\tabularnewline
&
$2_{ND}^{+}$&
91&
179$\pm$50&
1575&
1343\tabularnewline
$3_{\gamma}^{+}\rightarrow$&
$2_{ND}^{+}$&
28&
27$\pm$5&
2295&
2125\tabularnewline
$4_{\gamma}^{+}\rightarrow$&
$2_{ND}^{+}$&
20&
32$\pm$7&
2738&
2603\tabularnewline
&
$4_{ND}^{+}$&
49&
49$\pm$34&
1632&
1229\tabularnewline
\hline
\end{tabular}\end{center}
\end{table}

Finally, in table \ref{tab:out3} we have collected the out band transitions of the
low spin states of the ND $\gamma$ band. The experimental information is taken from
ref. \cite{toi}.
Our first concern  is to figure out to which extent our hypothesis of
low K-mixing is correct in the calculation. For that we compare the 
 $2_{\gamma}^{+}\rightarrow0_{ND}^{+}$ and the 
 $2_{ND}^{+}\rightarrow0_{ND}^{+}$ B(E2)'s, the former being twenty
times smaller. A similar reduction is found in the 
$4_{\gamma}^{+}\rightarrow2_{ND}^{+}$ $vs$ 
 $4_{ND}^{+}\rightarrow2_{ND}^{+}$ case, thus confirming the validity of our assumption.
The $\gamma$ energies compare quite well with the experimental results.
The B(E2)'s are in excellent agreement with the experiment in all but one case,
the corresponding to the  $2_{\gamma}^{+}\rightarrow0_{ND}^{+}$
transition that is severely under-predicted. Even so, the experimental value is three
time smaller than the equivalent transition inside the ND band.

\subsection{Other approaches}
\label{other}

 There has been two other recent attempts to explain theoretically
the structure of $^{40}$Ca such as we know it after the experimental
findings of Ideguchi {\it et al.} \cite{Ca40exp}  and  Chiara {\it et al.}
\cite{Ca40exp2}. Bender {\it et al.} \cite{Bender} approach the problem
in the framework of the constrained Hartree-Fock-BCS approximation. This is 
a calculation in the intrinsic frame, that is perfected by means of angular momentum and 
particle number projections. In a further step, the angular momentum 
projected solutions corresponding to different values of the deformation are
mixed using the Generator Coordinate Method. Modern Skyrme parametrizations 
 are used as effective interactions.
They produce a ND and a SD band with a span much larger than the experimental
one. The calculations are carried out only up to J=6 for the ND and J=10 for the
SD band.
 They assume an axially symmetric basis, therefore precluding
the description of the experimental $\gamma$ band.  Their 
ground state has a structure very 
similar to ours.
For the SD band their predicted transition quadrupole moments are quite 
large, but the agreement with the data is incomplete. No calculation of
out-band transitions is reported. There is a discussion of the np-nh
structure of the states that we find difficult to translate into our language.
They claim that the ND and SD bands do not have a well defined np-nh structure
contrary to what our results show. However, 
our percentages of np-nh components cannot be compared with
theirs directly.
 The second one applies the techniques of Antisymmetrized Molecular
Dynamics \cite{kanada}. They use an schematic interaction of Gaussian type.
The calculation produces many excited bands, and, in particular, the
three bands that we have studied in this paper. As we do, they obtain
a superdeformed band of 8p-8h nature. Their energetics is not
very satisfactory. The ND band-head, that is  experimentally the first excited state,
is calculated about 5~MeV too high, while the lowest 0$^+$ pertains to an oblate side
band not present in the experimental data. A bunch of states of unknown  
nature (J=2, 4, 6, and 8) are even lower. The SD band is also high but lower
than the ND band. The splittings in the SD band look healthy, but the ND
band does not resembles at all to a band. The $\gamma$ band appears also at very
high energy. Concerning the transition probabilities they are reasonable for the
ND band and rather small for the SD band. No out-band transitions are reported.

\section{Conclusions}
\label{sec:conc}

 In this article, we have shown that large scale shell model calculations
in a valence space comprising two major oscillator shells, can describe
the coexistence of a doubly magic ground state and deformed and superdeformed bands
in $^{40}$Ca. We have explored the role of the spherical mean field and the
correlations in bringing the n-particle n-hole configurations at very
low excitation energy. We have examined the mechanisms of mixing between
different np-nh configurations, and have analyzed
the structure of the physical states in terms of these configurations.
We find that the normally deformed bands are dominated by 4p-4h
configurations with important mixing of 6p-6h components. The superdeformed band
is clearly dominated by the 8p-8h configurations with small 6p-6h and 
4p-4h contributions. We submit that for J=8 the ND and SD bands
are accidentally degenerated and mix at 50\%. The ND intrinsic structure
is triaxial and produces a $\gamma$ band. In the ground state, the
doubly magic configuration amounts to 65\%, mixed mainly with 2p-2h states.
The comparison of the theoretical predictions with the experimental
data for the energetics of the three above mentioned bands is excellent.
We have computed also the electromagnetic transition
probabilities for in-band and out-band transitions. With some exceptions, the
agreement is also very good. In addition to the challenge of accounting
for super-deformation in a spherical shell model basis, {\it i. e.}
in the laboratory frame, our calculations and modeling support the
experimental claims on the existence of superdeformed bands at low
excitation energy in this mass region, even in doubly magic  $^{40}$Ca.

\acknowledgments{This work is supported by the  DGI(MEC)-Spain with a grant 
ref. BFM2003-1153, and by the IN2P3-France, CICyT-Spain collaboration agreements}

\end{document}